# Laser acceleration of protons from near critical density targets for application to radiation therapy.

Stepan S. Bulanov[1,2], Dale William Litzenberg[3], Alexander S. Pirozhkov[4], Alexander G. R. Thomas[1], Louise Willingale[1], Karl Krushelnick[1] and Anatoly Maksimchuk[1]

[1]FOCUS Center and Center for Ultrafast Optical Science, University of Michigan, Ann Arbor, Michigan 48109, USA

[2]Institute of Theoretical and Experimental Physics, Moscow 117218, Russia

[3]Department of Radiation Oncology, University of Michigan, Ann Arbor, Michigan 48109, USA

[4]Advanced Photon Research Center, Japan Atomic Energy Agency, 8-1-7 Umemidai, Kizugawa-shi, Kyoto 619-0215, Japan

Abstract

Laser accelerated protons can be a complimentary source for treatment of oncological diseases to the existing hadron therapy facilities. We demonstrate how the protons, accelerated from near-critical density plasmas by laser pulses having relatively small power, reach energies which may be of interest for medical applications. When an intense laser pulse interacts with near-critical density plasma it makes a channel both in the electron and then in the ion density. The propagation of a laser pulse through such a self-generated channel is connected with the acceleration of electrons in the wake of a laser pulse and generation of strong moving electric and magnetic fields in the propagation channel. Upon exiting the plasma the magnetic field generates a quasi-static electric field that accelerates and collimates ions from a thin filament formed in the propagation channel. Two-dimensional Particle-in-Cell simulations show that a 100 TW laser pulse tightly focused on a near-critical density target is able to accelerate protons up to energy of 250 MeV. Scaling laws and optimal conditions for proton acceleration are established considering the energy depletion of the laser pulse.

PACS : 52.38.Kd, 29.25.Ni, 52.65.Rr,

## I. Introduction

Laser acceleration of ions from both solid and gaseous targets attracts a lot of attention currently [1,2,3,4,5,6]. This is due to the recent development of laser systems capable of generating ultra-short pulses in the multiterawatt or even petawatt power range at high repetition rate. These systems are potentially able to accelerate ions to the energy of hundreds of MeV as has been shown by two dimensional (2D) and three dimensional (3D) particle-in-cell (PIC) computer simulations [7,8,9,10,11,12,13]. The laser accelerated charged particles can potentially be used for many applications including hadron therapy and other medical applications [14,15,16,17]. Though the highest ion energies yet experimentally achieved have been with single-shot multijoule picosecond pulse duration lasers (for example see Refs. [4,18]), ultra-short pulse (tens of femtoseconds) lasers may prove to be more advantageous for various applications and fundamental studies. This is due to the fact that high repetition rate is an important requirement for a number of applications including hadron therapy. Such lasers also are capable of delivering pulses of ultrahigh intensity [19], which is crucial for the investigation of new regimes of laser-matter interactions, especially for those which lead to proton acceleration.

It is repeatedly emphasized in the literature that the laser accelerated protons can be a practical source for treatment of oncological diseases, being complimentary to existing hadron therapy facilities [14,15,16]. Hadron therapy is a constituent part of radiation therapy, which makes use not only of high-energy ion beams but also of electron beams, x-rays and gamma radiation to irradiate cancer tumors (for details see Refs. [20,21] and the literature cited therein). Proton therapy has a number of advantages, since one of the main challenges of radiation therapy is to deliver a desired dose to the tumor without damaging the healthy tissues around the tumor. A proton beam is insignificantly scattered by atomic electrons and the range of protons ($g/cm^2$) with a given energy is fixed, which helps to avoid the undesired damaging of healthy tissues around and behind the tumor. The presence of a sharp maximum of proton energy loss in tissues (Bragg peak) provides a substantial increase in the radiation dose in the vicinity of the beam stopping point (see [20,21]). Up to the present time conventional particle accelerators have been used to produce proton beams with the required parameters. The use of laser accelerators seems to be very promising because of their compactness and potentially much lower cost. However the utilization of laser accelerated proton beam for the hadron therapy will require the acceleration of protons to the energy of 200-250 MeV. Aside from the high particle energy the therapeutic proton beam should provide a flux that is $\geq 10^{10}$ $s^{-1}$ with low energy spread of about 1%.

Since the small energy divergence of the proton beam is one of the main requirements for the laser accelerated protons to be used in hadron therapy several methods were proposed to generate such beams. One of the most promising ones is the use of double-layer (high Z/low Z) targets. It was theoretically proposed in Ref. [16], and the feasibility of such target design was verified experimentally [22,23]. Such target design is also critical in other regimes of laser proton acceleration such as the Directed Coulomb Explosion regime [24,25]. Most of the regimes of laser proton acceleration, such as mentioned above, are connected with the laser pulse interaction with thin and ultra-thin foils of solid density. In this paper we study the regime that utilizes targets with thickness several times greater then the laser pulse length and density of about critical [26,27,28,29,30].

The regime of laser ion acceleration from near critical density targets (NCD regime) is realized when the laser pulse propagates through a near critical density target, that is much longer then the pulse itself and the pulse forms a density channel. A portion of the electrons are accelerated in the direction of laser pulse propagation by the longitudinal electric field. The motion of these electrons generates a magnetic field, which circulates in the channel around the propagation axis. The region where the magnetic field is present follows the pulse. Upon exiting the channel, the magnetic field expands into the vacuum and the electron current is dissipated. This field has the form of a dipole in 2D and a toroidal vortex in 3D. The magnetic field displaces the electron component of plasma with regard to the ion component and a strong quasi-static electric field is generated that can both accelerate and collimate the ions. The accelerated ions originate from the thin ion filament that is formed along the axis of the propagation channel (see Fig. 1). In the case of long pulses the acceleration of helium ions up to 40 MeV from underdense plasmas was observed on the VULCAN laser [5]. The scaling and optimal conditions for this regime of acceleration were established in Ref. [31].

In what follows we focus on the NCD regime, which is potentially a more efficient mechanism of ion acceleration, and on the possibility of obtaining proton beams with energies that are of interest for medical applications. We show that a 100 TW laser pulse is able to produce protons with the maximum energy up to 250 MeV from a near critical density targets.

## II. Materials and methods.

In this section we describe the regime of proton acceleration form near critical density targets, the scalings of the optimal target thickness, radius of the laser generated channel and proton energy as well as the 2D PIC simulations of an intense laser pulse interaction with near-critical density targets. The principle scheme of laser proton acceleration is shown in Fig. 1.

### II.A. The optimal thickness and maximum proton energy.

We study the dependence of ion maximum energy on the target density and thickness, as well as on the focusing and power of the laser pulse in order to optimize the acceleration process. We should note here that the maximum proton energy is obtained for slightly overcritical density (with respect to nonrelativistic plasma density) targets and since the laser pulse should be able to establish a channel that goes through the target, the laser should be tightly focused on the front surface of the target to ensure penetration through the target. We show that it is not only important that the pulse penetrates the target, but also that it does not break into filaments. The latter will immediately reduce the effectiveness of acceleration. It is therefore necessary to establish matching between the dimensions of the focal spot, the position of the focus relative to the target boundary and the diameter of the self-focusing channel for each target density and thickness in order to avoid filamentation, as shown in Ref. [32]. The effectiveness of this mechanism depends on the efficient transfer of laser pulse energy into the energy of fast electrons that are accelerated along the propagation channel.

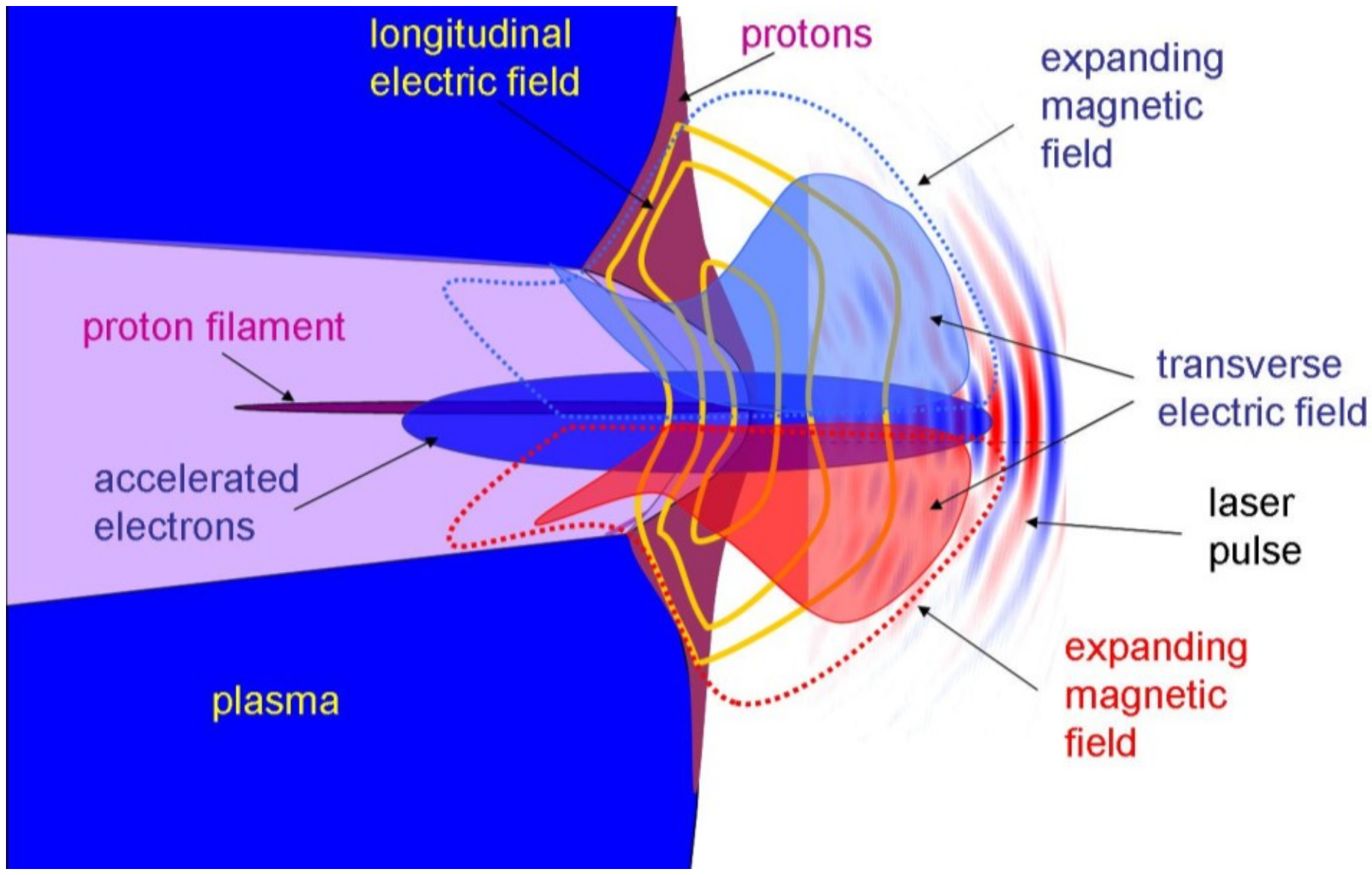

**Fig. 1. (color on-line)** The principal scheme of the acceleration mechanism

Since the laser pulse is interacting with the plasma of near critical density it is plausible to expect that the walls of the self-generated plasma channel will have density much higher than on-axis. Due to this fact the laser pulse will be confined inside the channel and the EM field of the pulse will not be able to penetrate the channel walls. As shown in Ref. [31], in this case the propagation of the laser pulse in the near critical density plasma can be approximated by the propagation of an EM wave in a wave guide [33]. The energy of the pulse in the waveguide is [31]

$$W_p = \pi R^2 \tau a^2 m_e c n_{cr} K , \qquad (1)$$

where $K = \sqrt{\pi/32}\left(J_1(\kappa R)^2 - J_0(\kappa R)J_2(\kappa R)\right)$ , $J_i$ is a Bessel function of the i-th order, $\kappa = 1.84/R$ , where $R = a^{1/2}(\omega/\omega_{pe})\lambda/2$ is the channel radius, $\tau$ is the pulse duration and $a$ is the maximum value of the dimensionless amplitude of the laser pulse EM field inside the waveguide.

It is easy to find a relationship between the dimensionless amplitude $a$ of laser beam inside the self-focusing channel of radius $R$ and the laser power, $P = W/\tau$, which is given by

$$a = \left[(8/\pi^2 K)(P/P_c)(n_e/n_{cr})\right]^{1/3}, \tag{2}$$

where $P_c = 2m_e^2 c^5/e^2$=17 GW. For $P = 100$ TW and $n_e/n_{cr}$=1 it yields $a \approx 40$.

Using (14) and the expression for the channel radius, we arrive at an important condition, which relates the channel radius, plasma density and laser power to each other

$$\frac{n_e R^3}{P^{1/2}} = \left(\frac{8}{\pi^2 K}\right)\frac{n_{cr}\lambda^3}{P_c^{1/2}} . \tag{3}$$

For a given value of laser power, we obtain the following dependencies of channel radius, $R$, and plasma density, $n_e$, on each other

$$R = \frac{\lambda}{2}\left(\frac{8}{\pi^2 K}\frac{P}{P_c}\right)^{1/6}\left(\frac{n_{cr}}{n_e}\right)^{1/3} \text{ or } n_e = n_{cr}\left(\frac{1}{8\pi^2 K}\frac{P}{P_c}\right)^{1/2}\left(\frac{\lambda}{R}\right)^3 . \tag{4}$$

These dependencies indicate that in order for the pulse to be able to establish a channel and propagate inside it, the radius of the focal spot should be less then *R*. Or, if the laser pulse focal spot radius is equal to *R*, the plasma density should be less than $n_e$ from Eq. (16). In Fig. 2 we present a series of curves, $R(n_e)$, for different values of laser pulse power. The values of ($R, n_e$) that lie below the curve should lead to the channel formation. As the power of the laser pulse increases more values of ($R, n_e$) become accessible. However for high densities, the area in the ($R, n_e$) plane is limited by the fact that different regimes of laser-plasma interaction that come into play, and the analysis presented here may become inapplicable.

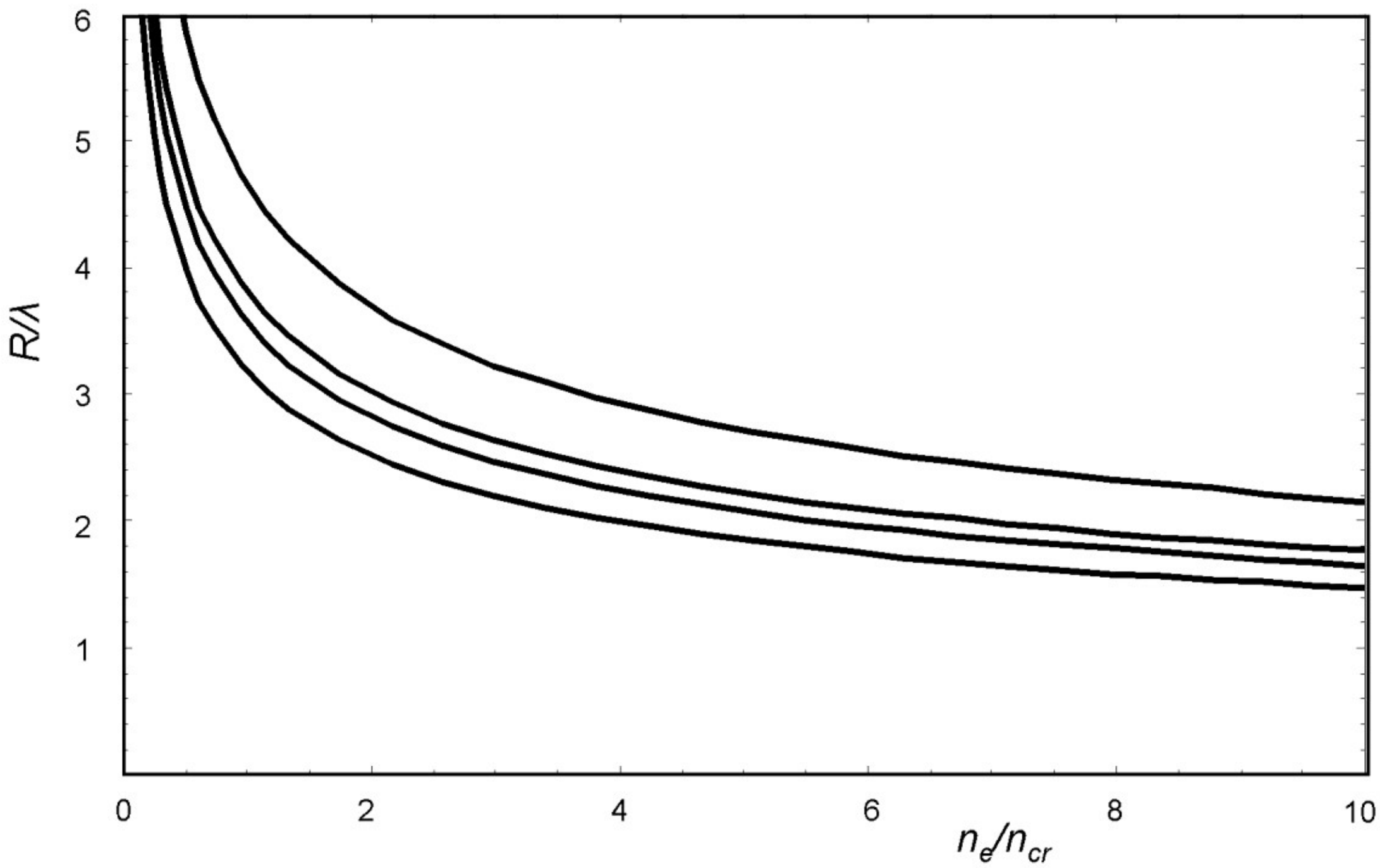

**Fig.2.** The dependence of the radius of the channel on plasma density for different values of laser pulse power (100 TW, 200 TW, 300 TW, 1 PW – curves from bottom to top).

After the laser pulse establishes the channel, the pulse propagates inside the self-generated channel through the target. Some of the electrons are accelerated in the forward direction in the regime that is similar to the blowout regime of electron acceleration in conventional laser wakefield acceleration (LWFA). These accelerated electrons generate a magnetic field. The region with magnetic field present moves behind the laser pulse. Upon the laser pulse exiting the channel, the magnetic field, which also exits the channel, begins to expand in the transverse direction along the target back surface. In doing so the magnetic field generates a quasistatic electric field that accelerates protons from the thin filament, formed along the propagation axis by the electron current.

The strength of the magnetic field and thus the strength of the quasistatic electric field, which accelerates the protons, depends on the energy of the electrons accelerated in the forward direction. So as long as the electrons are accelerated, the magnetic field will grow. That is why the optimum target thickness should be equal to the electron acceleration length. For targets of such density, as considered in this paper, the acceleration length is determined by the laser depletion length, i.e. the distance that the laser pulse can travel in plasma before transferring all its energy into plasma. This means that the target thickness should be equal to the laser depletion length to ensure optimal acceleration of protons. We can reformulate this condition in terms of the laser pulse energy transferred into the energy of electrons as follows. The optimal target thickness can be estimated from the requirement that of all the laser energy ($W_p$) to be transferred to the energy of electrons ($W_e$), which were initially in the volume of the propagation channel: $W_p = W_e$, where $W_e = \pi R^2 L_{ch} n_e a m_e c^2$, $L_{ch}$ is the channel length. The length of the channel is equal to the thickness of the target. Here, we assumed that the electrons acquire an average energy of $a m_e c^2$ after being pushed out from the channel in the transverse direction. We also assume that the thickness of the target is much larger than the pulse length, so the channel can be established. Then from Ref [31]

$$a = \frac{1}{K}\frac{n_e}{n_{cr}}\frac{L_{ch}}{L_p}, \tag{5}$$

where $L_p = \tau / c$ is the length of the laser pulse. If we express $a$ in terms of laser pulse energy then

$$\frac{n_e^{2/3} L_{ch}}{W_p^{1/3}} = C, \tag{6}$$

where the constant $C$ is

$$C = \left( \frac{4K^2 n_{cr} L_p^2}{\pi \lambda^2 m_e c^2} \right)^{1/3} . \qquad (7)$$

From this condition and from the condition of channel formation (15) we can obtain a scaling for plasma thickness with channel radius for optimal proton acceleration:

$$L_{ch} = L_p \frac{R^2}{\lambda^2} C_1 , \qquad (8)$$

where

$$C_1 = 4K \left( \frac{2\pi P_c}{\lambda^2 m_e c^3 n_{cr}} \right) . \qquad (9)$$

The maximum energy of the accelerated protons can be estimated from the fact that the acceleration itself takes place in the region which is of the order of the channel diameter. This assumption is justified by the fact that as the expansion of the electric fields at the rear of the target evolves and the region af expansion exceeds the channel diameter the efficiency of acceleration is greatly reduced. The electric field, at the rear of the target, should be of the order of the magnetic field generated by the accelerated electron bunch, i.e. $E \sim B \sim eN_e c$, where $N_e$ is the total number of electrons in the bunch. It was mentioned above that the laser acceleration of electrons inside the self-created channel is similar to the blow-out regime of acceleration achieved in LWFA. Then the number of electrons in the bunch can be estimated as $N_e \sim P^{1/2}$ [34]. Then the electric field also scales as the square root of the laser pulse power. The energy of protons accelerated by this field over the distance equal to the channel radius scales as

$$E_p \sim P^{2/3} \qquad (17)$$

## II. B. 2D PIC simulations

The simulations were performed using the REMP (Relativistic ElectroMagnetic Particle) code [35]. Space and time are measured in units of laser pulse wavelength, λ, and wave period, T=2π/ω, correspondingly, where ω is the laser pulse frequency. The grid mesh spacing is λ/40, and the time step is T/80. The total number of particles in the simulation box is about 5·106. A laser pulse with Gaussian temporal and spatial profiles is introduced at the left boundary. The pulse duration is τ=30 fs or 10λ/c (at 1/e in terms of the field amplitude). We used different focusing geometries from f/D=1.5 to f/D=4 and varied the position of the focus with regard to the left boundary so, that the laser is always focused at the front of the target. The target is composed of fully ionized hydrogen. The density is measured in units of the critical density, , me is the electron mass and e is the unit charge.

## III. The results of 2D PIC simulations.

In this section, we present the results of 2D PIC simulations of an intense laser pulse interaction with a near critical density target. As was mentioned in the previous section the laser pulse, tightly focused at the front of the target, makes a channel first in the electron and then in the ion density (Figs. 3a and 3b). Some fraction of the electrons are accelerated along the laser propagation direction. Upon the pulse exiting the plasma from the back, the electron current is dissolved (Fig. 3c,e). Approximately at that time the thin ion filament is formed along the central axis of the channel (Fig. 3d). As the interaction evolves, the electrons on the back surface are pushed inside the target (Fig. 3e) and the protons from the thin filament are accelerated in the forward direction (Fig. 3f).

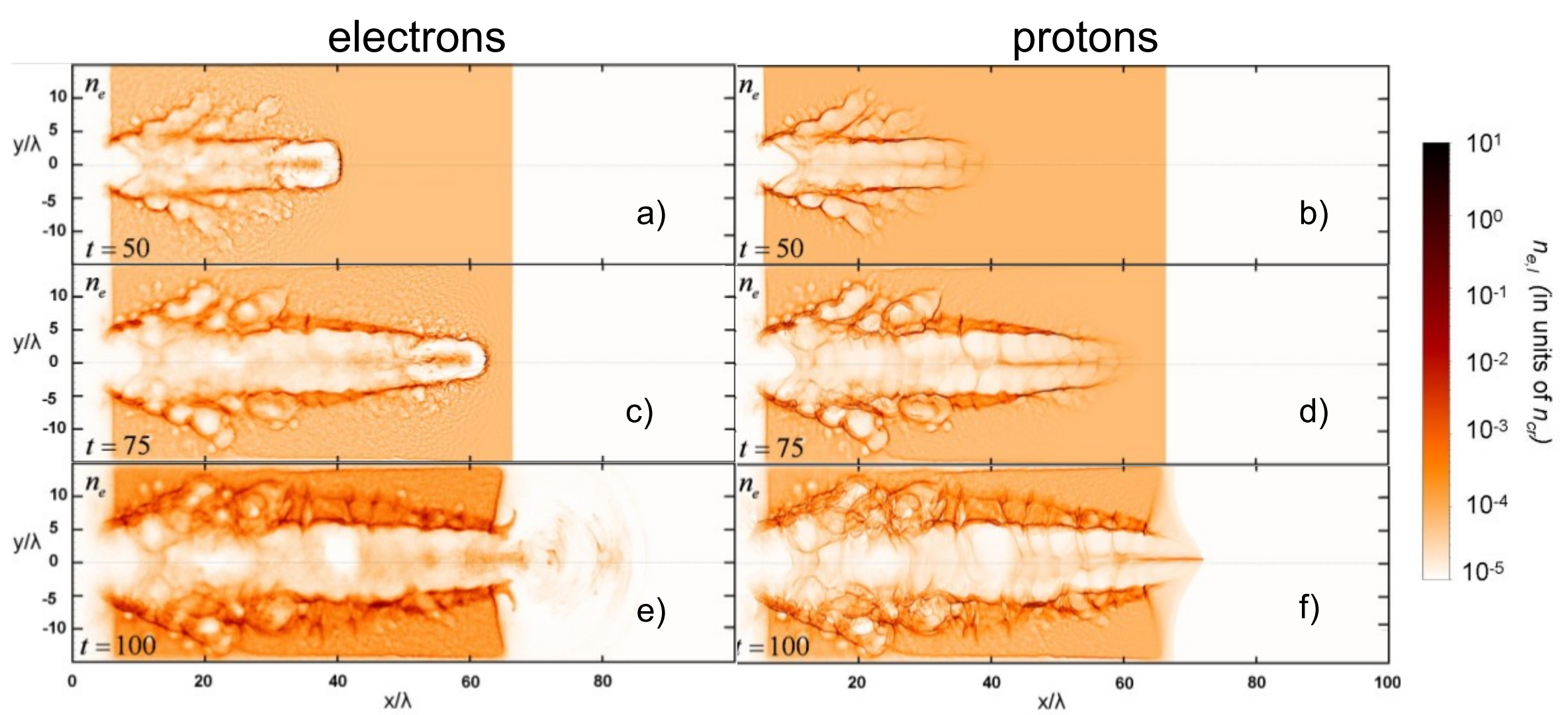

**Fig. 3. (color on-line)** The electron (a,c,e) and ion (b,d,f) density at different moments of time in a 100 TW laser pulse interaction with a $1n_{cr}$ 60 λ thick target.

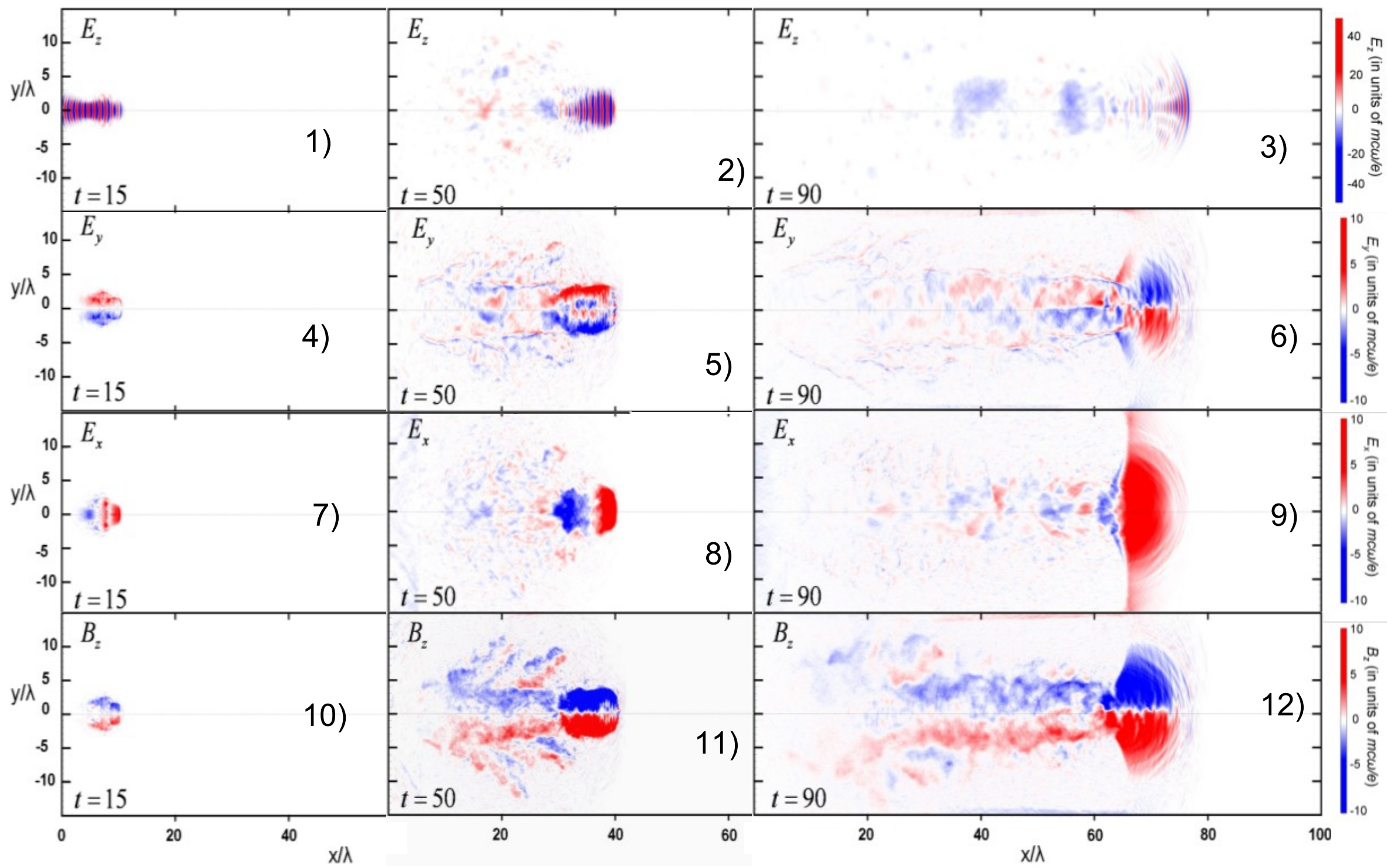

**Fig. 4 (color on-line)** The evolution of electric and magnetic fields inside the channel for a 100 TW laser pulse interaction with a $1n_{cr}$ 60 λ thick target.

The results of 2D PIC simulations enable us to identify the mechanism of acceleration, which can be illustrated by the evolution of the electric and magnetic fields of the pulse and those that are generated by the pulse inside the channel. The pulse is tightly focused at the front of the target (Fig. 4.1) in order to ensure the penetration through the near critical density plasma (Figs. 4.2 and 4.3). As the pulse propagates inside the channel the longitudinal field is formed (Figs. 4.7 and 4.8), which accelerates the electrons in the forward direction along the channel central axis (Figs 3a and 3c). This electron current generates a strong magnetic field (Figs. 4.10 and 4.11). When the laser pulse exits the plasma (Fig. 4.3) the magnetic field expands in the transverse direction (fig. 4.12). This field pushes the electrons back into plasma, thus displacing the electron component with regard to the ion component. Due to this the longitudinal charge separation, an electric field is established at the back of the target (Fig. 4.9). The charge separation field accelerates ions from the thin ion filament formed along the axis of the propagation channel (Fig. 3d). The collimating electric field in the transverse direction is due to the pinching of the electron beam in the magnetic field. The high density of on-axis electrons leads to the collimation of ions in the channel. We should also note here that in comparing Figs. 4.1 and 4.3 we see that the energy of the laser pulse is almost all depleted due to the energy transfer to plasma electrons. Thus the acceleration length of electrons is limited by the laser depletion length. This exact property of the interaction was used in the previous section to obtain the scaling for optimal acceleration conditions.

The evolution of the transverse component of the electric field (Figs. 4.4, 4.5 and 4.6) lets us track the formation of the channel and the appearance of the electron current. Such structure of the transverse electric field leads to the formation of thin proton filament along the propagation axis and to distinct features in the proton spectrum. Since we have here two acceleration mechanisms, the spectrum should have two temperatures. These mechanisms are: the acceleration of protons from the channel in the transverse direction and the acceleration of protons in the longitudinal direction from the thin filament. We will return to this feature when discussing the proton spectrum below. We should also note here that the transverse electric field generated by the electron current does not disappear when the pulse exits the target, but stays at the target back together with the magnetic and longitudinal electric fields, also expanding in the transverse direction (Fig. 4.6). This leads to the collimation of the accelerated proton beam, when it exits the channel.

In order to illustrate the acceleration of electrons and ions in the channel and at the exit from it, the evolution of electron and ion distribution in phase space $(x,p_x)$ is shown in Fig. 5. We can clearly see the electron current following the main pulse (Figs. 5a and 5c) and how it dissolves after the pulse exits the plasma (Fig. 5e) largely in the form of a return current. The formation of a thin ion filament along the propagation axis is demonstrated in Figs. 5b and 5d. After the electron current is dissolved and the longitudinal electric field is established at the back of the target the ions from this filament are accelerated in the forward direction (Fig. 5f).

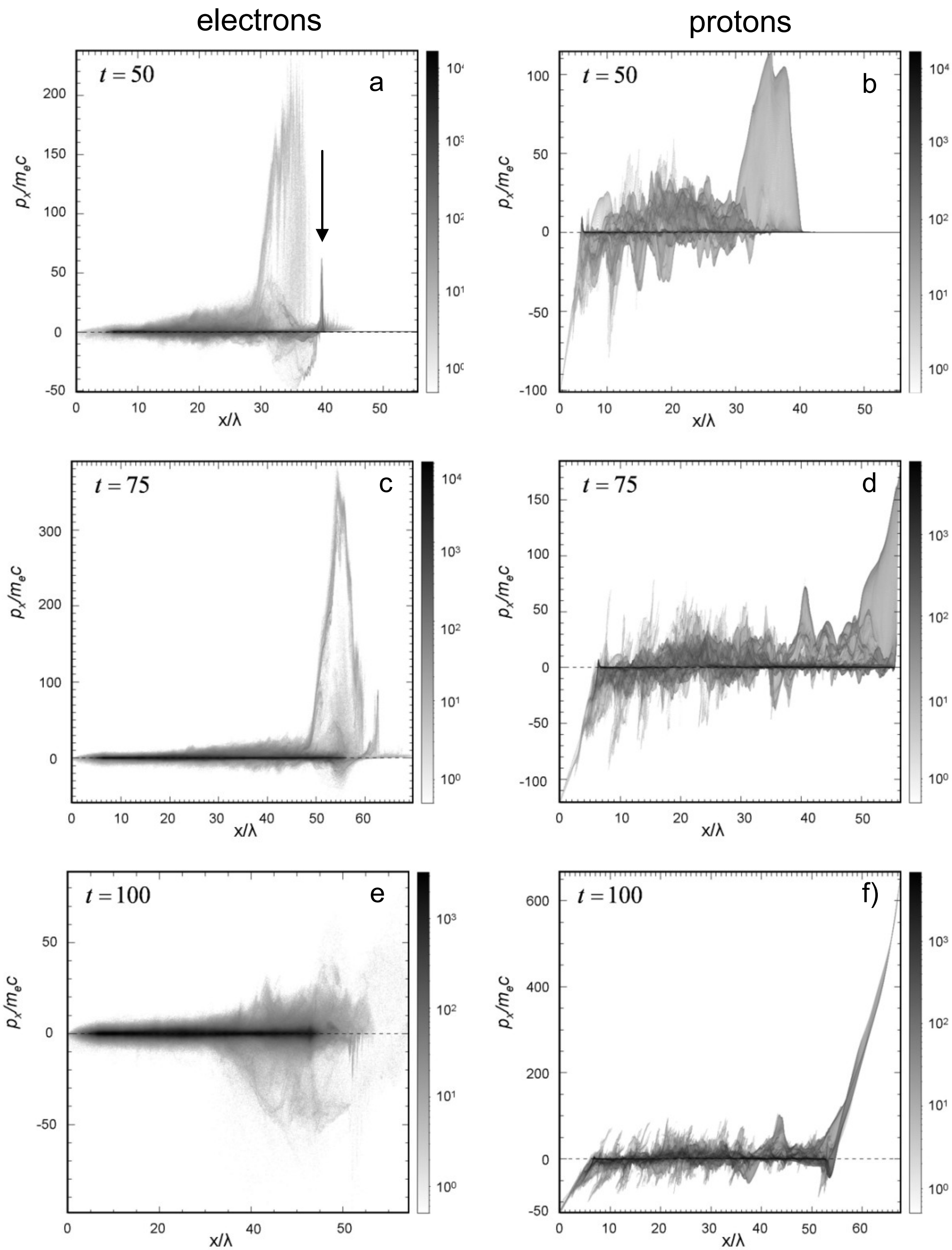

**Fig. 5** The evolution of the distribution of electrons (a,c,e) and ions (b,d,f) in the phase space ($x,p_x$) or a 100 TW laser pulse interaction with a 1$n_{cr}$ 60 λ thick target. The arrow in a) indicated the position of the laser pulse front.

This mechanism is possibly more effective than those that employ solid density ultra-thin foils as targets. The protons accelerated from a $1n_{cr}$ 60 λ thick target by a 100 TW laser pulse tightly focused (f/D=1.5) at the front of the target will have a maximum energy of 260 MeV (Fig. 6). The proton spectrum is shown in Fig. 6. The number of protons in the bunch with some energy spread of $\Delta E_p / E_p$ for $E_p = 100$ MeV is only three times larger then for $E_p = 250$ MeV. Due to this property of the accelerated proton spectrum, it would be possible to magnetically separate portions of the spectrum to give a tunable energy beam [36]. Also other selection schemes like simultaneous focusing and energy selection of proton beams with the use of radial, transient electric fields triggered on the inner walls of a hollow microcylinder by an intense subpicosecond laser pulse [37], the rotation of the proton beam in the longitudinal phase space with the use of the RF electric field, which is phase-adjusted with the pulse laser [38] or the transporting and focusing of proton beams by permanent magnet miniature quadrupole lenses [39] .

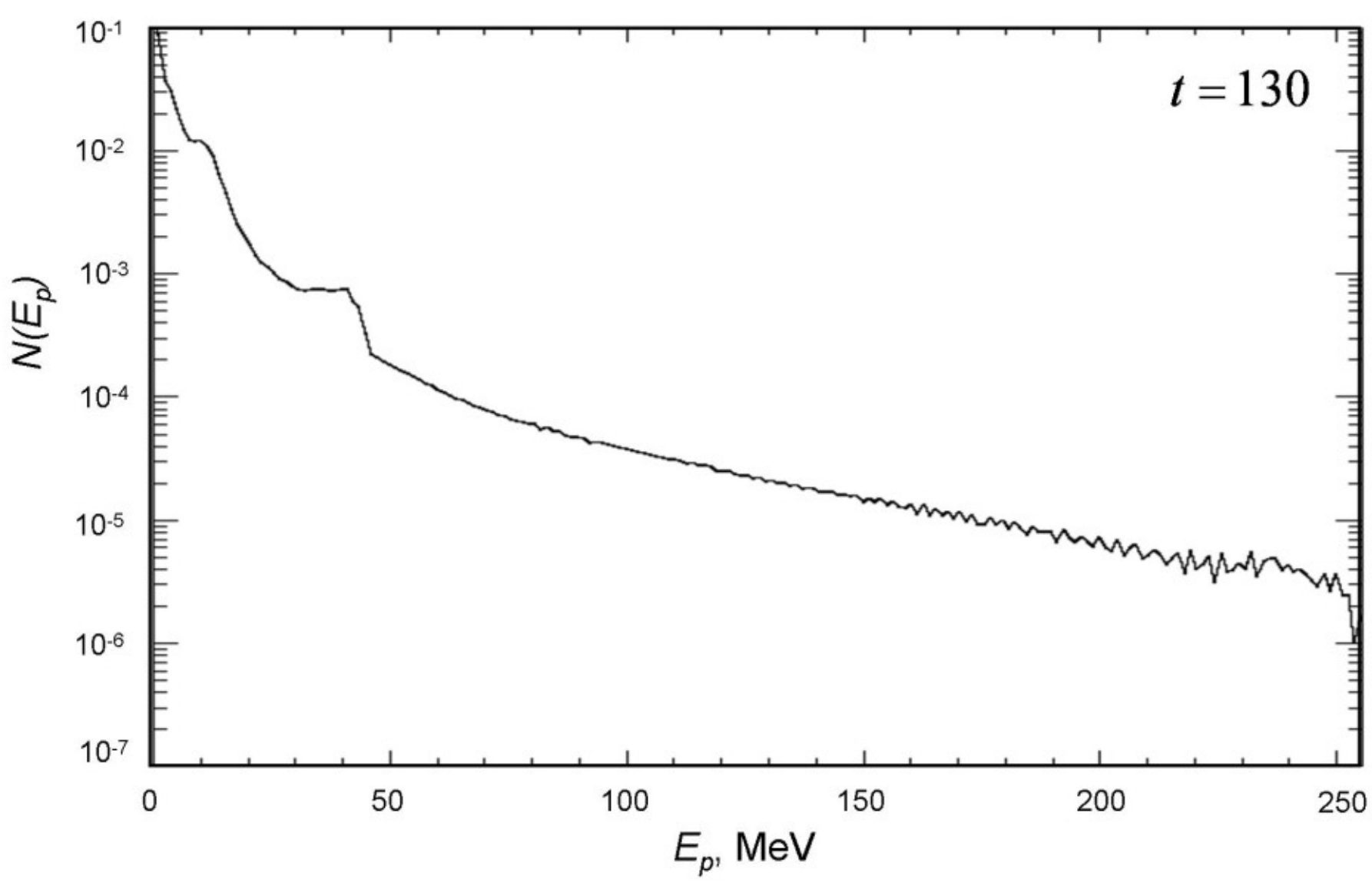

**Fig. 6** The spectrum of protons accelerated from $1n_{cr}$ 60 λ thick target by a 100 TW laser pulse.

The formation of the thin proton filament and the acceleration of protons from it can also be illustrated by the distribution of protons in the ($p_x, p_y$) plane (Fig. 7). The most energetic protons are generated along the laser propagation axis. However the spectrum of these protons does not have a monoenergetic structure. It rather expands over all possible proton energies. It is due to the fact that the accelerated protons emerge from the thin filament whose length is of the order of laser pulse length. That is why these protons experience different accelerating fields which result into such broad spectrum of protons.

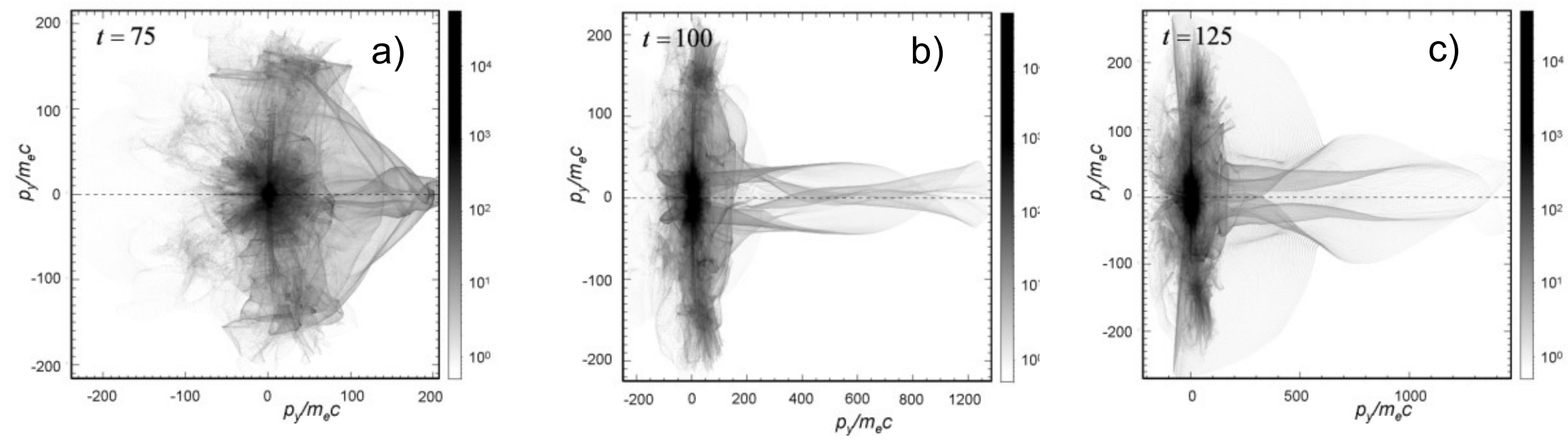

**Fig 7.** The distribution of protons in $(p_x,p_y)$ plane for a 100 TW laser pulse interacting with a $1n_{cr}$ 60 λ thick target for three time instances.

In what follows we present the scaling of the maximum proton energy with the laser pulse power obtained in 2D PIC simulations. In Fig. 8 we show the dependence of maximum proton energy on laser power for optimal target thickness for targets, i.e. for the value of target thickness which maximizes the proton energy for fixed laser parameters and target density (the laser has $f/D = 1.5$ and the target density is $n_e = 3n_{cr}$). The results of 2D PIC simulations for different combinations of focusing and target density $(f/D, n_e) = (1.5, 1n_{cr}), (1.5, 3n_{cr}), (3, 1n_{cr}), (3, 3n_{cr})$, and $(4, 1n_{cr})$ with optimal target thickness indicate the energy dependence on laser power of the form: $E_p \sim P^a$, where *a=0.7-0.8*. If we rewrite the scaling for proton energy obtained in the previous section for the 3D case in the 2D case, it will read as $E_p \sim P^{4/5}$ instead of $E_p \sim P^{2/3}$. It is in good agreement with the results of 2D PIC simulations. The lower value of *a* that follows from the simulations is due to the laser pulse energy loss because of the filamentation in the wings of the pulse (see Figs. 3 and 4), which reduces the maximum energy.

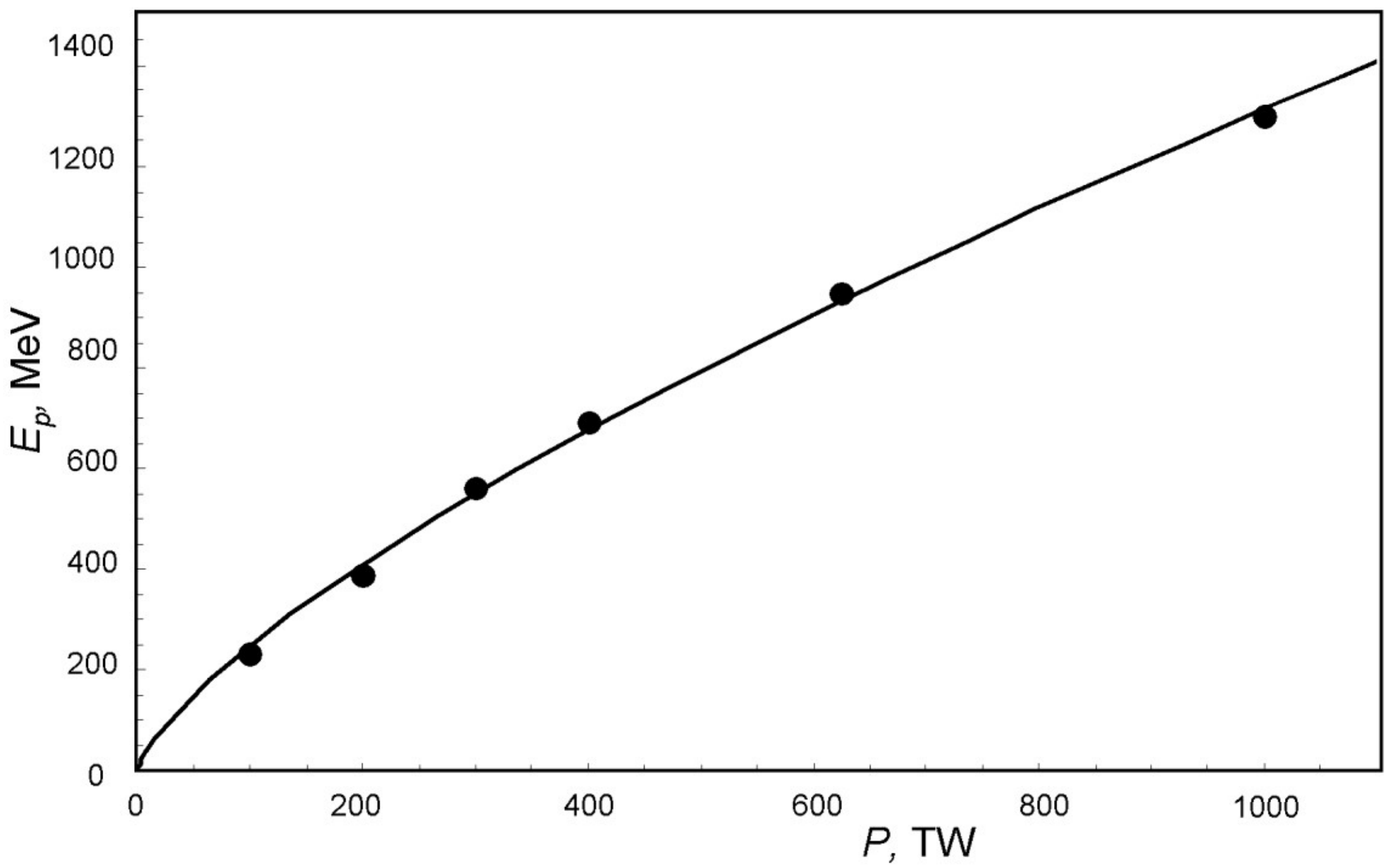

**Fig. 8.** The dependence of the proton maximum energy on the laser pulse power for $n_e = 3n_{cr}$, $f / D = 1.5$. The black circles are the results of 2D PIC simulations. The curve is the fit of the data by function $bP^a$. The thickness of the target was chosen to maximize the proton energy for the given laser pulse parameters and density of the target.

## IV. Discussions and Conclusions

In the present paper we were able to identify the properties of the laser-target interaction most favorable for proton acceleration from near critical density targets for medial applications. We utilized the scaling for the optimal acceleration conditions derived in Ref. [31]. We established a relation between the radius of the laser generated plasma channel and the density of the undisturbed plasma as a condition of the channel creation. This allowed us to set a limit on the radius of the focal spot that will lead to the channel formation for a given density and laser pulse power:

$$R = \frac{\lambda}{2}\left(\frac{8}{\pi^2 K}\frac{P}{P_c}\right)^{1/6}\left(\frac{n_{cr}}{n_e}\right)^{1/3}$$

or this can be rewritten in the form analogous to the scaling in [31]:

$$\frac{n_e R^3}{P^{1/2}} = \left(\frac{8}{\pi^2 K}\right)\frac{n_{cr}\lambda^3}{P_c^{1/2}}$$

If this scaling is combined with the scaling for optimal target thickness, then a relation between the channel radius and optimal target thickness can be established:

$$L_{ch} = L_p \frac{R^2}{\lambda^2} C_1, \text{ where } C_1 = 4K\left(\frac{2\pi P_c}{\lambda^2 m_e c^3 n_{cr}}\right),$$

i.e. the target thickness scales as the square of the channel radius. We also established a scaling for the maximum proton energy with the laser pulse power:

$$E_p \sim P^{2/3}$$

as the result of the magnetic and electric fields estimates at the rear of the target.

These scalings were used to guide 2D PIC simulations of the laser pulse interaction with near critical density targets. The results of the simulations indicate that this regime of acceleration leads to a significant reduction of the laser pulse power necessary to produce proton beams with the energy of interest for medical applications compared to other schemes of proton acceleration.

Moreover the spectrum of protons has a characteristic two temperature behavior, which indicates two mechanisms of acceleration. The low energy part of the spectrum corresponds to the protons accelerated in the transverse direction by the electric fields in the channel. The high temperature part corresponds to the protons accelerated in the forward direction from the thin dense filament formed along the laser propagation axis inside the channel. Though the spectrum has an exponential structure, the characteristic high temperature allows for the extraction of a fraction of the proton beam with small energy divergence and still maintaining a significant number of protons. We showed that the scaling of maximum proton energy with laser power, established in 2D PIC simulations, follows the 2D version of the power law obtained analytically.

## Acknowledgements

This work was supported by the National Science Foundation through the Frontiers in Optical and Coherent Ultrafast Science Center at the University of Michigan. The authors would like to thank Dr. T. Zh. Esirkepov for providing the REMP code for simulations.